\documentclass[journal]{IEEEtran}
\usepackage{amsmath,amsfonts}
\usepackage{algorithmic}
\usepackage{array}
\usepackage{ragged2e} 
\usepackage[caption=false,font=normalsize,labelfont=sf,textfont=sf]{subfig}
\usepackage{textcomp}
\usepackage{stfloats}
\usepackage{booktabs}  
\usepackage{url}
\usepackage{verbatim}
\usepackage{graphicx}
\usepackage{xcolor}
\usepackage[ruled,linesnumbered]{algorithm2e}
\def\BibTeX{{\rm B\kern-.05em{\sc i\kern-.025em b}\kern-.08em
    T\kern-.1667em\lower.7ex\hbox{E}\kern-.125emX}}
\usepackage{balance}
\begin{document}
\title{Robustness of One-to-Many Interdependent Higher-order Networks Against Cascading Failures}
\author{ Cheng Qian, Dandan Zhao, Bo Zhang, Ming Zhong, 
 Jianmin Han, Shenghong Li, Hao Peng, Wei Wang

\thanks {This work was supported by the Key Project of the
National Natural Science Foundation of China under grant no. 62536007, the Zhejiang Province Science Foundation under grant nos. LD24F020002, LQ24F020025, LZYQ25F020003, the Zhejiang Province's 2025 "Leading Goose + X" Science and Technology Plan under grant no.2025C02034, the Program for Youth Innovation in Future Medicine, Chongqing Medical University, No. W0150.

Cheng Qian, Dandan Zhao, Ming Zhong and Jianmin Han are with the School of Computer Science and Technology, Zhejiang Normal University, Jinhua 321004, China.  

Bo Zhang and Shenghong Li are with the School of Cyber Science and Engineering, Shanghai Jiao Tong University, Shanghai 200240, China.

Hao Peng is with the Key Laboratory of Intelligent Education Technology and Application of Zhejiang Province and the School of Computer Science and Technology, Zhejiang Normal University, Jinhua 321004, China. Email: hpeng@zjnu.edu.cn.

Wei Wang is with the School of Public Health, Chongqing Medical University, Chongqing, 400016, China. Email: wwzqbx@hotmail.com and wwzqbc@cqmu.edu.cn.

Hao Peng and Wei Wang are co-corresponding authors.
} } 

\markboth{IEEE TRANSACTIONS ON RELIABILITY,~Vol.~*, No.~*, *~2025}%
{How to Use the IEEEtran \LaTeX \ Templates}

\maketitle

\begin{abstract}
In the real world, the stable operation of a network is usually inseparable from the mutual support of other networks. In such an interdependent network, a node in one layer may depend on multiple nodes in another layer, forming a complex one-to-many dependency relationship. Meanwhile, there may also be higher-order interactions between multiple nodes within a layer, which increases the connectivity within the layer. Inter-layer dependencies and intra-layer connectivity may become key factors affecting network reliability, because failures within a layer will propagate to another layer through dependencies, and the cascading effects within and between layers may trigger catastrophic network collapse. However, existing research on one-to-many interdependence often neglects intra-layer higher-order structures and lacks a unified theoretical framework for inter-layer dependencies. Moreover, current research on interdependent higher-order networks typically assumes idealized one-to-one inter-layer dependencies, which does not reflect the complexity of real-world systems. These limitations hinder a comprehensive understanding of how such networks withstand failures. Therefore, this paper investigates the robustness of one-to-many interdependent higher-order networks under random attacks. Depending on whether node survival requires at least one dependency edge or multiple dependency edges, we propose four inter-layer interdependency conditions and analyze the network's robustness after cascading failures induced by random attacks. Using percolation theory, we establish a unified theoretical framework that reveals how higher-order interaction structures within intra-layers and inter-layer coupling parameters affect network reliability and system resilience. Additionally, we extend our study to partially interdependent hypergraphs. We validate our theoretical analysis on both synthetic and real-data-based interdependent hypergraphs, offering insights into the optimization of network design for enhanced reliability.
\end{abstract}


\section{Introduction}
\IEEEPARstart{T}{here} is increasing evidence that there are a large number of networks in the real world that are composed of various relationships \cite{2010Networks,cohen2010complex,bianconi2018multilayer}, such as social networks composed of social relationships \cite{9603289}, biological networks composed of predator-prey relationships \cite{liu2020robustness}, and control networks supported by control mechanisms \cite{zhang2016cascading,li2023review}. These networks often cannot maintain normal functional operations independently and usually require other networks to provide resources and support, showing obvious interdependence \cite{shahrivar2017spectral,tang2025network}. Compared with isolated networks \cite{dong2015system}, the interdependence between networks will undoubtedly affect the robustness of the network \cite{rosato2008modelling,alessandro2010complex,zhang2019robustness}. This is because when a node fails or is attacked, the connectivity between nodes within the layer and the dependencies between nodes between layers will trigger cascading failures \cite{Dynamic,cellai2013percolation}, eventually leading to the collapse of the entire network \cite{zhang2018cascading}. In order to address the cascading failure risks in interdependent networks, a lot of modeling and robustness analysis work has been carried out \cite{cao2021percolation,chattopadhyay2016estimation,chattopadhyay2017designing}. For example, Li and Zhang \cite{li2025cascading} studied the cascading failures of interdependent water supply and power networks under random failures. Kays et al. \cite{kays2025modeling} studied the simulation of flood propagation and cascading failures in interdependent transportation and stormwater networks. Zhou et al. \cite{zhou2023combined} studied the reliability and resilience of dependent network systems under mixed cascading failures.

It is worth noting that there are some typical cases in the real world that highlight the system vulnerability caused by the high dependence between critical infrastructure systems. For example, the large-scale power outage in Italy on September 28, 2003 is a typical case of cascading failures. The accident was caused by a failure in the power transmission lines, which led to the shutdown of the power plant and subsequently affected the operation of the communication network. As a result, multiple critical nodes in the communication network failed, and the collapse of the communication network in turn hindered the recovery of the power network, ultimately causing the entire system to collapse. To deeply understand such failure mechanisms, Buldyrev et al. \cite{buldyrev2010catastrophic} modeled the power network and the communication network as a one-to-one interdependent network, revealing how faults propagate between networks and eventually cause the entire system to collapse. On this basis, Gao et al. \cite{gao2012robustness} further studied the robustness of n interdependent networks. Havlin et al. \cite{havlin2015percolation} systematically explored the robustness of the network of networks.

However, since the one-to-one dependency relationship may be too idealistic in reality, researchers have proposed a variety of more general models \cite{dong2012percolation, huang2011robustness, PhysRevE.97.032306, radicchi2015percolation}. For example, the partial interdependence model allows some nodes to survive without relying on another layer of the network; in the one-to-many dependency model, a node can rely on multiple nodes in another layer of the network. Parshani et al. \cite{parshani2010interdependent} discovered the phase transition from first-order to second-order percolation in the interdependent network by reducing the inter-layer coupling strength. Zhou et al. \cite{zhou2013percolation} studied the robustness of partially interdependent scale-free networks; Shao et al. \cite{shao2011cascade} analyzed the cascading failure behavior that may occur in coupled networks with multiple supporting dependencies. Dong et al. studied the robustness of n interdependent networks with partially supporting dependencies \cite{dong2013robustness}, the robustness of interdependent networks with many-to-many dependencies\cite{dong2019robustness}, and the robustness of coupled networks with multiple effective dependencies \cite{dong2021percolation}. In addition, Zhang et al. \cite{zhang2020asymmetric} focused on asymmetric interdependent networks with multiple dependencies, and Zheng et al. \cite{zheng2022robustness} studied the robustness of circularly interdependent networks. Han and Yi \cite{wei2019percolation} proposed a conditional dependency group model, which allows dependent nodes to function normally when the failure rate of nodes in the dependency group does not exceed the tolerance $\gamma$, thereby enhancing the fault tolerance of the system.

Although the above improved model is closer to reality to a certain extent, it still only considers pairwise interactions between nodes within its layer, ignoring the possible higher-order interactions between multiple nodes. With the development of network science, hypergraphs \cite{ghoshal2009random,bretto2013hypergraph,6980103}, as one of the typical representations of higher-order networks, have received extensive attention and research \cite{peng2022disintegrate,peng2024message}. Compared with graphs, a hypergraph consists of nodes and hyperedges \cite{PhysRevLett.85.5468,battiston2020networks,latifi2008robustness,8091121}, where nodes represent individuals or elements in the system, and hyperedges capture higher-order relations or collective interactions among multiple nodes. This structure not only overcomes the limitation of edges connecting only two nodes but also provides a more effective way to characterize higher-order interactions among multiple nodes \cite{wan2022multilayer,majhi2022dynamics}. For example, in a power network, a node can represent a power generation device, and a hyperedge corresponds to a power station; in a communication network, a node is a communication device, and a hyperedge represents a control center. In recent years, hypergraph structures have been considered to be introduced into interdependent network modeling to characterize higher-order interactions between nodes \cite{peng2025robustness,liu2023threshold}. Sun et al. \cite{sun2021higher} constructed a general framework to analyze the critical behavior of higher-order percolation in multi-layer hypergraphs and revealed the impact of structural correlation on robustness. Chen et al. \cite{chen2024catastrophic} proposed a threshold model to analyze the cascading failure in interdependent hypergraphs. Our previous work also showed that the robustness of interdependent hypergraphs is more fragile than that of isolated hypergraphs\cite{qian2024cascading,QIAN2025110497}. However, existing studies have primarily focused on one-to-one interdependence structures, while the more realistic one-to-many interdependent mechanisms remain largely unexplored.

This paper focuses on the modelling and analysis of cascading failures on inter-layer one-to-many interdependency hypergraph networks that are more in line with practical scenarios. The aim is to evaluate the robustness of the network in the event of node failures \cite{zio2011modeling,9190058}. Since nodes have multiple dependency edges, we can naturally consider four different dependency scenarios: i) Node survival depends on all edges, for example, when assembling a computer, the monitor, host, keyboard, mouse and other hardware must be complete and indispensable; ii) Node survival requires at least one dependency edge, for example, the computer only needs at least one socket to provide power; iii) Node survival allows at most $\left\lfloor K\gamma \right\rfloor$ dependency edges to fail (using floor to handle non-integers), for example, a distributed computing system with $K$ redundant servers can still function correctly if no more than $\gamma$ proportion of the servers fail; and iv) Node survival allows at most $M$ dependency edges to fail, for example, to ensure the normal operation of the computer system, at most $M$ key servers or network nodes can be tolerated to fail in order to maintain overall network connectivity and service stability. Based on the above four inter-layer interdependence conditions, we use generating functions and self-consistent equations to accurately calculate the size of the network's giant component (GC) after random failures through a unified theoretical framework. Our analysis reveals how the intra-layer higher-order interaction structure, inter-layer dependencies, and coupling strength jointly affect the network's reliability and the system's resilience.

The main contributions of this paper are as follows:

\begin{itemize}
\item{We propose a one-to-many interdependent hypergraph network model and analyze the dynamics of node cascading failure by randomly attacking nodes in the network. The study shows that node survival rate, intra-layer connectivity, inter-layer dependency, and coupling strength significantly impact network robustness}.
		
\item{In nodes with multiple dependent edges, we further explore the impact of the failure of a dependent node in another layer on the node and its impact mechanism. We propose four conditional dependencies based on the inter-layer dependency and describe them using a unified theoretical analysis framework.}

\item{Considering that there may be autonomous nodes in real-world networks (i.e. nodes that can exist or function independently without relying on other nodes), we believe that the survival condition of at least one surviving dependent edge may be too strict. Therefore, we propose a one-to-many partial dependency hypergraph network model in this context.}

\end{itemize}

This paper is organized as follows. Section II will describe the proposed system model in detail and list all parameter definitions in this paper in a table. Section III provides an in-depth theoretical analysis of the four inter-layer interdependency conditions in the system model. Section IV presents experimental validation of the theoretical analysis using both numerical simulations and empirical tests. Finally, Section V summarizes this paper's main work and research contributions and looks forward to future research directions.

\section{SYSTEM MODEL}
This section will describe our proposed one-to-many interdependent hypergraph network model in detail and list the initial parameters involved in Table \ref{table}.

\begin{table}[htbp]
    \centering
    \caption{Initial Parameters}
    \renewcommand{\arraystretch}{1.3} 
    \setlength{\tabcolsep}{6pt} 
    \begin{tabular}{|>{\centering\arraybackslash}m{2cm}|>{\justifying\arraybackslash}m{6cm}|}
        \hline
        \textbf{Symbol} & \textbf{Description} \\ \hline
        $N$ & The total number of nodes in the network layer \\ \hline
        $A$, $B$& Two layers of one-to-many interdependent hypergraph networks \\
				\hline
				$\hat{l}_{A}$, ($\hat{l}_{B}$)& 
                The probability that a node in the layer $A$ ($B$) is connected to a factor node in the GC along an edge\\ \hline
				$l_{A}, (l_{B})$ &	
                The probability that a factor node in the layer $A$ ($B$) is connected to a node in the GC along an edge  \\ \hline
			$q^{A}$, $(q^{B})$ & The percentage of nodes in network $A$ ($B$) that depend on network $B$ ($A$)  \\ \hline
				$k$, $m$ &The number of hyperdegrees and the hyperedge cardinality on hypergraphs \\ \hline
				${\langle k\rangle}, {\langle m\rangle}  $&The average hyperdegree and the average hyperedge cardinality on hypergraphs  \\ \hline
				$P(k), \hat{P}(m)$&The hyperdegree distribution and the hyperedge cardinality distribution on hypergraphs \\ \hline          $\lambda$&Power exponent in power-law distribution\\ \hline
                $k_{max}, k_{min}$&The maximum and minimum hyperdegrees in the hypergraph are set to $\sqrt{N}$ and 2\\ \hline
                $m_{max}, m_{min}$&The maximum and minimum hyperedge cardinality in the hypergraph are set to $\sqrt{N}$ and 2\\ \hline      
$ K$&The number of directed inter-layer dependency nodes (equivalent to the inter-layer out-degree of a node)\\ \hline           
${\langle K\rangle}$&The average number of directed inter-layer dependency nodes\\ \hline
$\tilde{P}(K)$&The degree distribution of the number of directed inter-layer dependency nodes\\  \hline
$\delta$&The maximum number of failed dependency edges that a node can tolerate while still remaining functional.\\  \hline
 \end{tabular}
      \label{table}
\end{table}

\begin{figure*}[!t]
		\centering
		\includegraphics[width=1.0\textwidth]{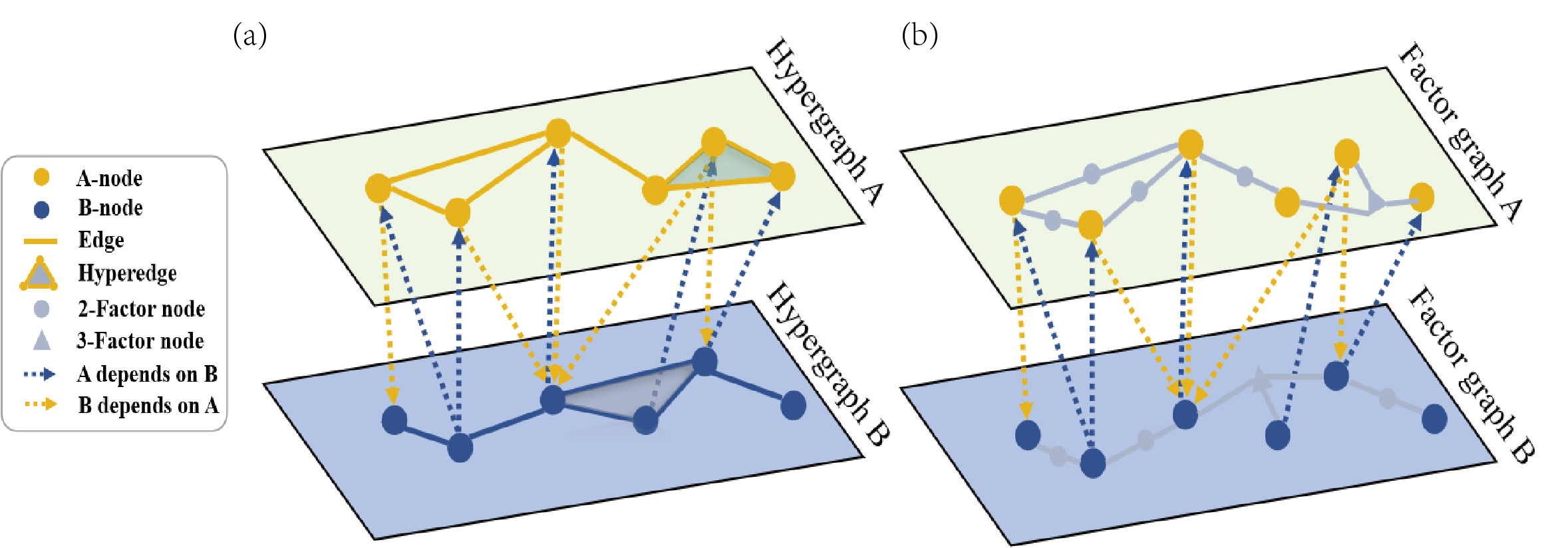}
		\caption{The interdependence hypergraph is transformed into an interdependence factor graph. Solid lines within layers denote intra-layer connectivity, while dashed arrows across layers represent unidirectional inter-layer dependency. An arrow from hypergraph $B$ pointing to hypergraph $A$ indicates that hypergraph $A$ depends on hypergraph $B$. (a) Hypergraphs $A$ and $B$ each consist of 6 nodes; hypergraph $A$ contains 5 hyperedges, and hypergraph $B$ contains 4 hyperedges (where an edge is equivalent to a hyperedge of cardinality 2). (b) The corresponding factor graph representation of (a), where each hyperedge is transformed into a factor node. Different polygon shapes represent hyperedges with different cardinalities.}
		\label{fig1}
\end{figure*}

\subsection{Intra-layer Higher-order Interactions}
This paper considers using hypergraphs to model higher-order interactions within a layer. Hypergraphs are composed of nodes and hyperedges. Their advantage is that they can express common interactions between multiple nodes through hyperedges, which goes beyond the traditional pairwise interactions between two nodes. The most critical parameters of a hypergraph are the node hyperdegree $k$ and the hyperedge cardinality $m$. Usually, the nodes within a hyperedge are considered to be fully connected. To give the simplest example, three hyperedges with a cardinality of 2 (equivalent to edges) or a hyperedge with a cardinality of 3 can form a triangular loop in the network, as shown in Fig. \ref{fig1}(a). Therefore, to identify higher-order interactions in the network, we usually convert the hypergraph into a factor graph: the nodes and factor nodes in the factor graph correspond to the nodes and hyperedges in the hypergraph, respectively \cite{sun2021higher}, as shown in Fig. \ref{fig1}(b). In our proposed model, a node is considered functional if it belongs to at least one functional hyperedge, while a hyperedge is deemed functional if it contains at least one functional node. A hyperedge is considered failed if and only if all the nodes it contains have failed.

\subsection{Inter-layer One-to-many Dependencies}
\begin{figure}[!t]
\includegraphics[width=0.5\textwidth]{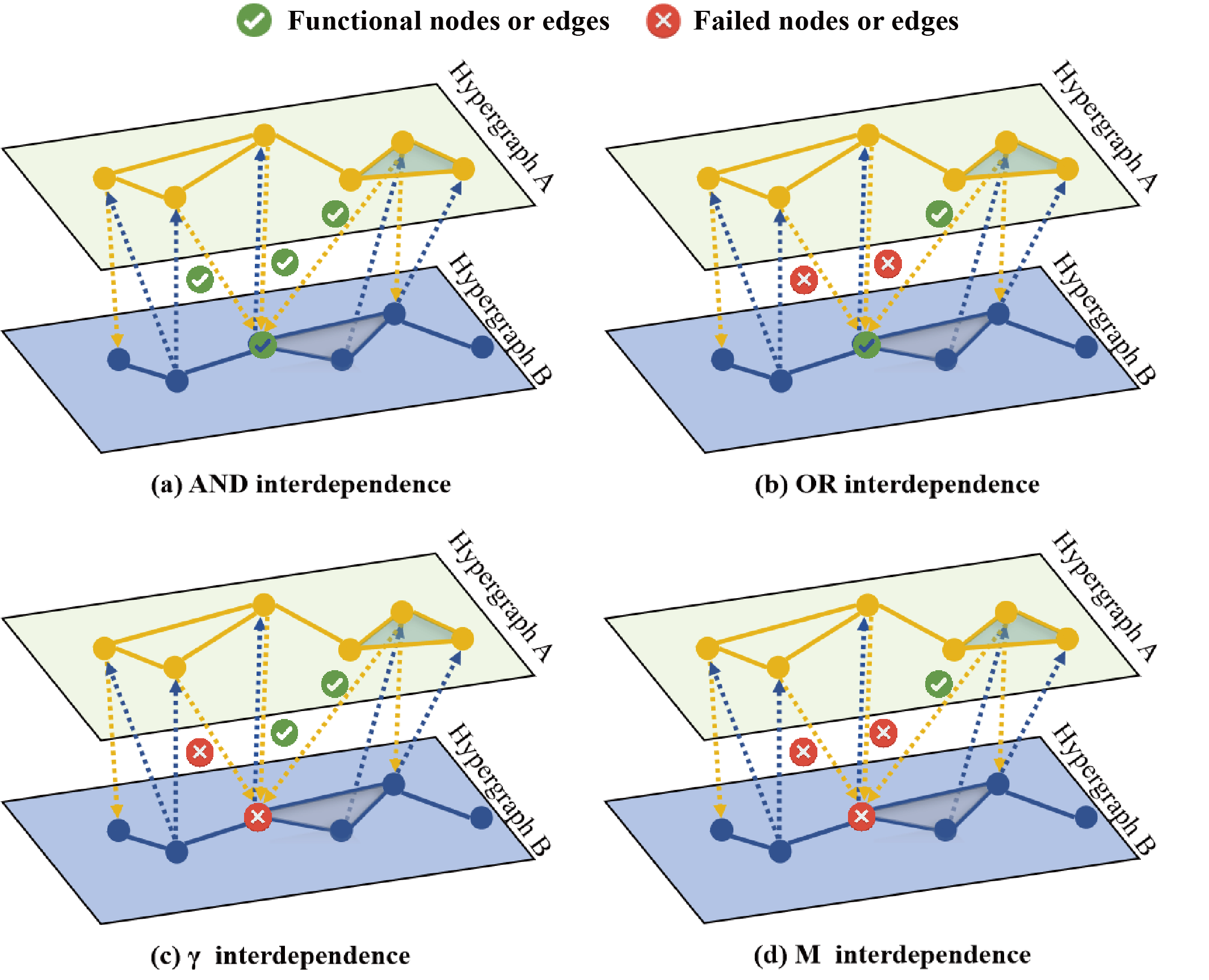}
\caption{Schematic diagrams of four conditional dependencies. Whether a node survives is determined by the survival of its dependent edges. (a) AND interdependence, a node survives if all its dependent edges survive; (b) OR interdependence, a node survives if at least one of its dependent edges survives; (c) $\gamma$ interdependence, where we assume $\gamma$ = 0.3, and a node fails if the number of failures among its dependent edges exceeds $\left \lfloor K\gamma \right \rfloor$; and (d) $M$ interdependence, where we assume $M$ = 1, and a node fails if the number of failures among its dependent edges exceeds $M$.}
\label{fig2}
\end{figure}

In our proposed interdependent hypergraph system, we consider the existence of one-to-many dependencies between nodes in layers \cite{dong2019robustness}, i.e., a node in hypergraph $A$ is provided with resources by multiple nodes in hypergraph $B$, and vice versa. However, this does not exclude one-to-one dependencies or cases where some nodes do not depend on nodes in another layer. This paper proposes four conditional dependencies for node survival: i) AND interdependence. The survival of a node requires the support of all dependent edges (if any). A typical example is a supply chain network, which cannot work if the raw material is unavailable; ii) OR interdependence, the survival of a node requires the support of at least one dependent edge. Compared with the strict constraint of one-to-one complete interdependence, this kind of dependent system will be more robust; iii) $\gamma$ interdependence, where $\gamma$ is the tolerance for failure ($\gamma \in [0,1]$). The survival of a node is related to its own $K$ value, and at most $\left\lfloor K\gamma \right\rfloor$ edges are allowed to fail; and iv) $M$ interdependence, the survival of a node is related to the $M$ value, and at most $M$ edges are allowed to fail. There are essential differences in the mechanisms of $\gamma$ interdependence and $M$ interdependence: $\gamma$ interdependence adopts a relative fault tolerance mechanism (proportional tolerance), whose fault tolerance is proportional to the node's dependency, and is suitable for modeling systems whose robustness increases with the dependency strength; while $M$ interdependence adopts an absolute fault tolerance mechanism (threshold tolerance), where each node can only tolerate the failure of at most $M$ dependent edges, and is suitable for systems with a fixed fault tolerance threshold. The schematic diagram of the above four conditional dependencies is shown in Fig. \ref{fig2}.

It is worth noting that whether a node survives in our model is related to the properties of the node itself. For example, if the node itself does not depend on another layer of nodes, then under the AND interdependence condition, it can still survive in the context of 0 dependent edges. Similarly, $\gamma$ interdependence ($K=0$) and $M$ interdependence ($0<M$), the node can still survive. Only when OR interdependence occurs does the node fail because it does not meet the support of at least one dependent edge.

\begin{figure}[!t]
\centering
\includegraphics[width=0.5\textwidth]{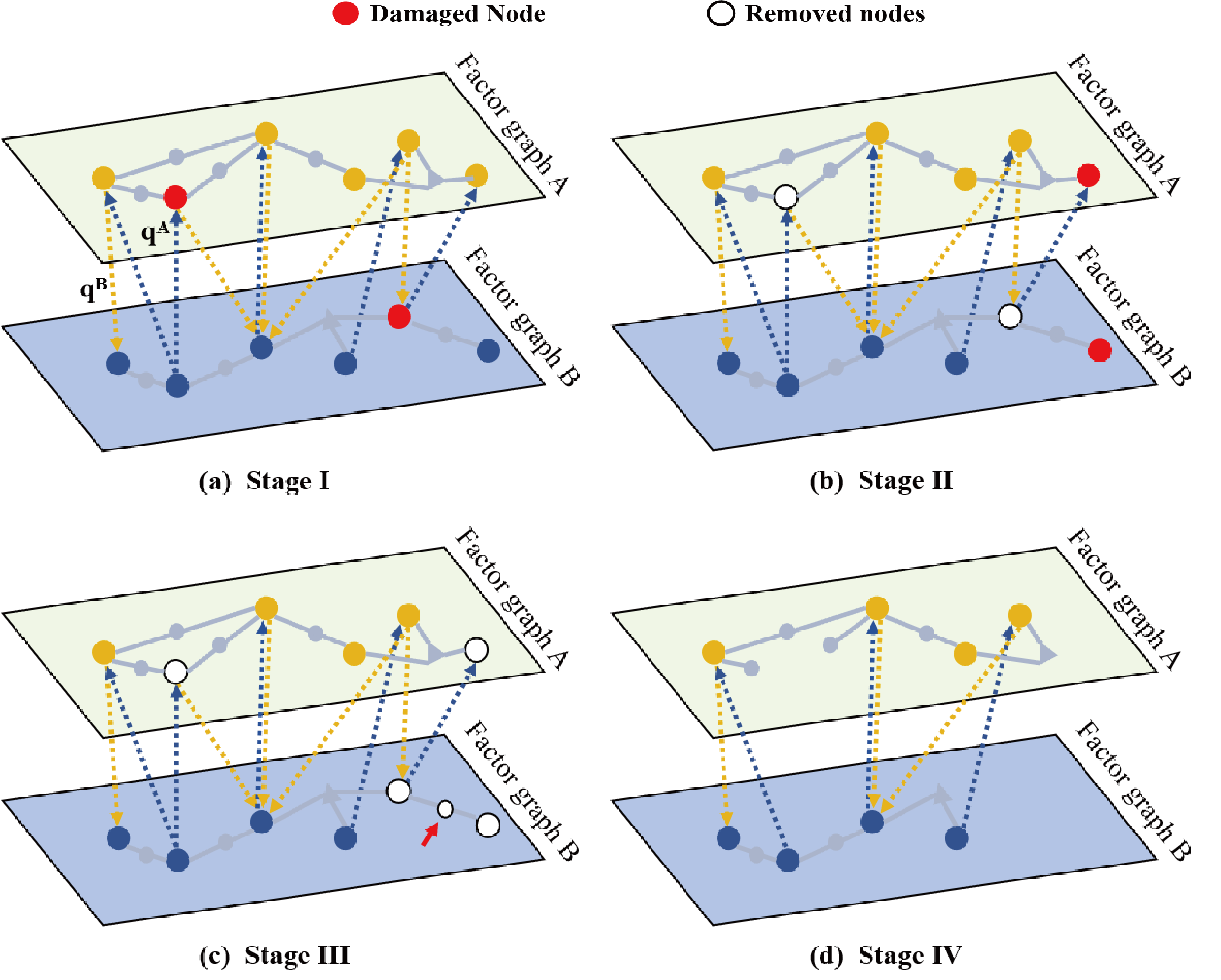}
\caption{Cascading failures on partially interdependent hypergraphs. For a more intuitive approach, we use the corresponding factor graph. (a) Stage I, nodes in networks $A$ and $B$ are respectively attacked by initial attacks and damaged, respectively, with a node damage rate $(1-p)=1/6$; (b) Stage II, the damaged nodes are removed from the network, and the fault further propagates within and between layers, resulting in damage to dependent node that do not satisfy OR interdependency and connected node that are not in GC; (c) Stage III, because all nodes connected to the factor node (hyperedge) are damaged, the factor node is damaged (indicated by the red arrow). (d) Stage IV, the network reaches a final stable state, and the remaining size represents the effective structure after cascading failures.}
\label{fig3}
\end{figure}

\subsection{Coupling Strengths Across Layers}
 Coupling strength describes the degree of interdependence between different layers in the system. The coupling strength level directly affects the system's complexity and robustness. Before this, we believed that nodes in OR interdependence must require resources from nodes in another layer; otherwise, they would fail. However, this condition is too strict in real life. Here, we consider that nodes can survive without relying on nodes in another layer. In a partially one-to-many interdependent hypergraph network, only a proportion of $q^{A}$ of nodes in network $A$ depends on nodes in network $B$, and vice versa, as shown in Fig. \ref{fig3}(a). By scaling the parameters $q^{A}$ and $q^{B}$, the coupling strength between layers can be changed. When $q^{A}$=$q^{B}$=0, the two layers of the network are independent of each other. That is, the failure of a node in network $A$ does not affect the robustness of network $B$. When $q^{A}$=$q^{B}$=1, the inter-layer dependency is consistent with the original OR interdependence model. This also shows that as the network coupling strength increases, the inter-layer cascade process gradually dominates the intra-layer cascade process, which significantly impacts the robustness of the entire system. After cascading failure occurs in a partially dependent network, the functional node's size is defined as a partially dependent giant component (PDGC).

\subsection{Cascading Failures Across Layers}
The cascading failure process of the macroscopic interdependent hypergraph network can essentially be seen as two microscopic dynamic processes: intra-layer cascade (breadth percolation) and inter-layer cascade (depth percolation). Initially, the node fails due to random attacks. During the breadth percolation process, initial failures may further lead to the failure of other nodes within the same layer due to the removal of intra-layer connectivity edges. In contrast, during the depth percolation process, failures may propagate across multiple layers through inter-layer dependency relationships. The survival of nodes belonging to GC needs to meet conditional dependencies, and nodes that do not belong to the {mutually connected giant component  (MCGC) will also fail. The robustness of the network can be reflected by analyzing the cascading failure of the network. When facing the same network damage, the fewer failed nodes in the network, the more robust it is.

We provide an example of a cascading failure on a partially interdependent hypergraph under the OR interdependence condition, as illustrated in Fig. \ref{fig3}. Hypergraphs $A$ and $B$ both contain 6 nodes, the inter-layer coupling strength $q^{A}=q^{B}=2/3$, and the survival probability of each layer node is $p=5/6$. After the inter-layer cascading failure and the intra-layer cascading failure occur, the total number of network nodes $N$ in the steady state is reduced from 12 to 8.

\section{THEORETICAL ANALYSIS}
This section studies how the interdependence structure affects the network robustness. We obtain exact theoretical solutions under different dependency conditions with the help of self-consistent equations \cite{feng2015simplified} and generating functions.

\subsection{GC in Single Hypergraphs}

First, consider the probability that a node in a layer belongs to the GC. We convert a hypergraph into a factor graph, as described above, to analyse this. In the factor graph representation, the node degree distribution is \(P(k)\), and the factor node degree distribution is \(\hat{P}(m)\).

Considering that the network is initially randomly attacked, $p$ represents the probability that the node is not damaged, and $\mu_{\infty}$ represents the size of the network in the final steady state after the cascading failure ends. We define the generating function $L_{0}(x)$ of the node degree distribution as
	\begin{equation}
	L_{0}(x)=\sum_{k} P(k) x^{k}.
	\label{1}
	\end{equation}
the generating function $L_{1}(x)$ of excess degree is
	\begin{equation}
	L_{1}(x)=L_{0}^{\prime}(x)/L_{0}^{\prime}(1)=   \sum_{k} P(k)k x^{k-1}/\left \langle k \right \rangle.
	\label{2}
	\end{equation}
    
Similarly, the generating functions $\hat{L}_{0}(x)$ and $\hat{L}_{1}(x)$ associated with the factor node are defined as follows:
	\begin{equation}
	\begin{aligned}
	\hat{L}_{0}(x)=\sum_{m}\hat{ P}(m) x^{m}, \\
	\hat{L}_{1}(x)=\hat{L}_{0}^{\prime}(x)/\hat{L}_{0}^{\prime}(1)=   \sum_{m} \hat{P}(m)m x^{m-1}/\left \langle m \right \rangle.
	\end{aligned}
	\label{3}
	\end{equation}

We introduce two variables \(l\) and \(\hat{l}\) to quantify the probability associated with the GC, which are defined in Table \ref{table}. If we want the node with degree \( k \) reached along the edge to belong to GC, we need to ensure that at least one of its remaining $(k-1)$ edges is connected to a factor node belonging to GC; similarly, if we want the factor node with degree \( m \) reached along the edge to belong to GC, we need to ensure that at least one of its remaining $(m-1)$ edges is connected to a node belonging to GC. When the probability of node survival is $p$, through the self-consistent equations of $l$ and $\hat{l}$, we can get

	\begin{equation}
	\begin{aligned}
	\hat{l}= \sum_{m} \hat{P}(m)m [1-(1-l)^{m-1}]/\left \langle m \right \rangle, \\
	 \quad l=  p\sum_{k}  P(k)k[1-(1-\hat{l})^{k-1}]/\left \langle k \right \rangle.
	\end{aligned}
	\label{4}
	\end{equation}

	For any node selected in the network, the probability that the node belongs to GC is the probability that at least one of its $k$ edges leads to GC. Therefore, we can finally calculate the size of the network in the steady state as 
	
	\begin{equation}
	 \mu_{\infty}=  p[1-  \sum_{k}P(k)(1-\hat{l})^k] .
	\label{5}
	\end{equation}
	
 The above can be transformed into the form of a generating function according to Eq. (\ref{1}):
	
\begin{equation}
\mu_{\infty} =p\left[1-L_{ 0}\left(1-\hat{l}\right)\right].
\label{6}
\end{equation}

In a single-layer hypergraph, only a second-order phase transition occurs. The critical threshold point \( p_{II}^* \), representing the maximum attack intensity the hypergraph can endure, is defined as
  
\begin{equation}
\begin{aligned}
p_{II}^{*}=\left \langle k \right \rangle/\left \langle k \left (k-1 \right ) \right \rangle  \cdot \left \langle m \right \rangle/  \left \langle m\left (m-1 \right ) \right \rangle .
\end{aligned}
\label{7}
\end{equation}

This paper mainly studies the robustness of the inter-layer structure of the network. It is not convenient to expand too much here. Interested readers can refer to Ref.\cite{sun2021higher}.

\subsection{MCGC in interdependent Hypergraphs}
In the real world, seeing hypergraph networks in isolation is difficult because a network usually requires other networks to provide resources. When a network is damaged, it will also affect the functions of its dependent networks. Connectivity causes cascading failures of nodes within a layer, while dependencies cause cascading failures of nodes between layers. Therefore, when an interdependent hypergraph network is attacked, cascading failures will occur both within and between layers, and this process will continue until no further nodes fail. We call the remaining functional node part of the network in the steady state the MCGC. The one-to-many interdependent hypergraph model proposed in this paper allows one node to depend on multiple nodes, and different dependency conditions will lead to different robustness of the network.

We define the probability function of satisfying the inter-layer dependency condition as $\chi_{i}(y)$, which represents the probability that a randomly selected node on the layer $i$ satisfies the dependency condition. For any surviving node selected in the interdependent network, the probability that it belongs to the MCGC is equal to the product of the probability that it belongs to the GC within its own layer and the probability that the inter-layer dependency satisfies the condition $\chi_{i}(y)$. We define the probability that a node in a hypergraph network $A$ belongs to MCGC as $\mu_{\infty}^{\mathrm{A}}$, and the probability that a node in a hypergraph network $B$ belongs to MCGC as $\mu_{\infty}^{\mathrm{B}}$. Therefore, we can deduce

\begin{equation}
\begin{aligned}
\mu_{\infty}^{\mathrm{A}}=p\left[1-L_{\mathrm{A} 0}\left(1-\hat{l}_{\mathrm{A}}\right)\right] \chi_{\mathrm{B}}\left(\mu_{\infty}^{\mathrm{B}}\right), \\
\mu_{\infty}^{\mathrm{B}}=p\left[1-L_{\mathrm{B} 0}\left(1-\hat{l}_{\mathrm{B}}\right)\right] \chi_{\mathrm{A}}\left(\mu_{\infty}^{\mathrm{A}}\right).
\end{aligned}
\label{8}
\end{equation}

where $\hat{l}_{\mathrm{A}}$ and $\hat{l}_{\mathrm{B}}$ satisfy  
\begin{equation}
\begin{aligned}
\hat{l}_{\mathrm{A}}= \ 1-\hat{L}_{\mathrm{A} 1}\left(1-l_{\mathrm{A}}\right) ,\\
l_{\mathrm{A}}=p\left[1-L _{\mathrm{A} 1}\left(1-\hat{l}_{\mathrm{A}}\right)\right]
\chi_{\mathrm{B}}\left(\mu_{\infty}^{\mathrm{B}}\right).\\
\hat{l}_{\mathrm{B}}= \ 1-\hat{L}_{\mathrm{B} 1}\left(1-l_{\mathrm{B}}\right) ,\\
l_{\mathrm{B}}=p\left[1-L _{\mathrm{B} 1}\left(1-\hat{l}_{\mathrm{B}}\right)\right]
\chi_{\mathrm{A}}\left(\mu_{\infty}^{\mathrm{A}}\right), \\
\end{aligned}
\label{9}
\end{equation}   
  
Next, we consider four different conditional dependencies. The innovation of this paper is that we propose a unified framework to obtain different dependency scenarios by changing the value of the parameter $\delta$ (i.e., the maximum number of dependency edges that are allowed to fail while the node remains functional). The inter-layer conditional dependency function is
\begin{equation}
\chi_i(y) = \sum_{K} \tilde{P}_i(K)  \sum_{S=0}^{\delta} \binom{K}{S} (1 - y)^S y^{K- S} .
\label{10}
\end{equation} 
  
Where $K$ is the out-degree of the node in layer $i$, that is the number of edges it depends on in another layer. $\tilde{P}_i(K)$ is the degree distribution of the out-degree of the node in layer $i$. The first summation term $\sum_{K}$ represents the sum of the probabilities of survival of nodes with different out-degrees; the second summation term $\sum_{S=0}^{\delta}$ represents the number of edges that are allowed to fail, ranging from 0 to $\delta$. The rest of the formula constitutes a binomial distribution, representing the probability that $(K-S)$ of the $K$ edges can edge to a surviving node in another layer. The following is how we set different parameter values $\delta$ to meet different dependency conditions.

\subsubsection{AND interdependence} 
AND interdependence requires that all $K$ dependency edges of a node are alive, meaning that all dependency nodes in the dependency network must also be functional. In this case, the parameter $\delta$ can be set to 0; at most, zero edges are allowed to fail. At this time, the dependency function $\chi_{i}^{_{A}}(y) $ can be further simplified to

\begin{equation}
\chi_{i}^{_{A}}(y) = \sum_{K} \tilde{P}_i(K) y^{K}.
\label{11}
\end{equation}  

At this point, it can be simplified into a standard generating function form. It is not difficult to see that the one-to-one interdependent hypergraph network is a special case of AND dependency when $\tilde{P}(1)$= 1 and the node $j$ in layer $B$ that node $i$ in layer $A$ depends on happens to also rely on node $i$ \cite{qian2024cascading}.

\subsubsection{OR interdependence} 
OR interdependence condition is not as strict as the AND interdependence condition. It only requires that at least one of the $K$ dependent edges of a node remain alive. In this case, the parameter $\delta$ can be set to $(K-1)$, which means that at most $(K-1)$ edges are allowed to fail (when $K=0$, the node fails). In this case, the dependency function $\chi_{i}^{_{O}}(y) $ can be further simplified to
\begin{equation}
\chi_{i}^{_{O}}(y) = 1-\sum_{K} \tilde{P}_i(K) (1-y)^{K}.
\label{12}
\end{equation}  


\subsubsection{$\gamma$ interdependence} 
The $\gamma$ interdependence considers the different abilities of different out-degree nodes to tolerate the failure of dependent nodes. The parameter $\gamma$ refers to the degree of failure allowed, also known as the tolerance. The $\gamma$ interdependence allows at most $\left\lfloor K\gamma \right\rfloor$ edges out of the $K$ dependent edges of a functional node to lose their function. In this case, the parameter $\delta$ can be set to $\left \lfloor  K \gamma \right \rfloor$, and nodes with different numbers of dependent edges have different $\delta$ values. At this time, the dependency function $\chi_{i}^{_{\gamma}}(y)$ is

\begin{equation}
\chi_{i}^{_{\gamma}}(y) = \sum_{K} \tilde{P}_i(K)  \sum_{S=0}^{\left \lfloor  K \gamma\right \rfloor} \binom{K }{S} (1 - y)^S y^{K- S}.
\label{13}
\end{equation} 

There are several interesting special cases here: when $\gamma=0$, the $\gamma$ interdependence can be converted into AND interdependence; when $\gamma\to 1$ ($\gamma \neq 1$), the $\gamma$ interdependence can be converted into OR interdependence; when $\gamma=1$, the dependency function can be simplified to be always equal to 1, which means that the dependency relationship in the interdependent hypergraph network no longer exists.

\subsubsection{M interdependence} 
The $M$ interdependence mainly considers that if the number of failed dependent edges of a node exceeds threshold $M$, the node will be removed. Repeat this process until the number of failed dependent edges of all nodes does not exceed $M$. In this case, the parameter $\delta$ can be set to $M$, and nodes with different numbers of dependent edges have the same $\delta$ value. At this time, the dependency function $\chi_{i}^{_{M}}(y)$ is

\begin{equation}
\chi_{i}^{_{M}}(y) = \sum_{K} \tilde{P}_i(K)  \sum_{S=0}^{M} \binom{K}{S} (1 - y)^S y^{K- S}.
\label{14}
\end{equation} 

Don’t worry about the superscript $M$ in the sum exceeding $K$, because according to the combinatorial property, when $M>K$, $\binom{K}{S}=0$; this means that the excess combinatorial terms are equal to zero, so these terms will not contribute anything to the sum.

Solving the critical threshold $p_{I}^{*}$ in a one-to-many interdependent network. Theoretically, according to Eqs. (\ref{8}) and (\ref{9}), we can express $\hat{l}_{\mathrm{A}}$ and $\hat{l}_{\mathrm{B}}$ as $\hat{l}_{\mathrm{A}}= \Gamma_{1} (p,\hat{l}_{\mathrm{B}})$, $\hat{l}_{\mathrm{B}}= \Gamma_{2}  (p,\hat{l}_{\mathrm{A}})$. If there is a first-order phase transition point in the one-to-many interdependent network, then the tangent lines of the two functions have a tangency point at $p_{I}^{*}$, that is,

\begin{equation}
\begin{aligned}
\partial \Gamma_{1}\left( p^{*}_{I}, \hat{l}_{\mathrm{B}}\right)/\partial \hat{l}_{\mathrm{B}} \cdot \partial \Gamma_{2}\left( p^{*}_{I}, \hat{l}_{\mathrm{A}}\right)/\partial \hat{l}_{\mathrm{A}}=1.
\end{aligned}
\label{15}
\end{equation}

\subsection{PDGC in Partially Dependent Hypergraphs}
The above conditional dependencies mainly consider the strength of the dependencies between nodes in different layers. Next, we propose partial dependencies to consider the coupling strength between nodes in different layers.

In a partially dependent network that satisfies OR interdependence, we allow nodes in a layer to survive without relying on nodes in another layer. Therefore, we define the probability that a node in network $A$ needs to rely on a node in network $B$ as $q^{A}$; similarly, the probability that a node in network $B$ needs to rely on a node in network $A$ is $q^{B}$. 

When a surviving node is randomly selected from a partially dependent network, the node belongs to PDGC if: i) the node belongs to GC in its layer; ii) if the node has inter-layer dependency, it also needs to rely on a functional node in another layer. We can derive the following:

\begin{equation}
\begin{aligned}
\mu_{\infty}^{\mathrm{A}}=p\left[1-L_{\mathrm{A} 0}\left(1-\hat{l}_{\mathrm{A}}\right)\right]\left[(1-q^{A}) + 
q^{A}\chi_{\mathrm{B}}\left(\mu_{\infty}^{\mathrm{B}}\right)\right],\\
\mu_{\infty}^{\mathrm{B}}=p\left[1-L_{\mathrm{B} 0}\left(1-\hat{l}_{\mathrm{B}}\right)\right]\left[(1-q^{B}) + 
q^{B}\chi_{\mathrm{A}}\left(\mu_{\infty}^{\mathrm{A}}\right)\right].
	\label{16}
    \end{aligned}
\end{equation}
where $\hat{l}_{\mathrm{A}}$ and $\hat{l}_{\mathrm{B}}$ satisfy  

	\begin{equation}
	\begin{aligned}
	\hat{l}_{\mathrm{A}}= 1-\hat{L}_{A1}(1-l_{\mathrm{A}}), \\
	l_{\mathrm{A}}=p  [1-{L}_{A1}(1-\hat{l}_{\mathrm{A}})] [(1-q^{A})+q^{A}\chi_{\mathrm{B}}\left(\mu_{\infty}^{\mathrm{B}}\right)].\\
   \hat{l}_{\mathrm{B}}= 1-\hat{L}_{B1}(1-l_{\mathrm{B}}), \\
	l_{\mathrm{B}}=p  [1-{L}_{B1}(1-\hat{l}_{\mathrm{B}})] [(1-q^{B})+q^{B}\chi_{\mathrm{A}}\left(\mu_{\infty}^{\mathrm{A}}\right)].\\
	\end{aligned}
	\label{17}
	\end{equation}

It is not difficult to see from the above formula that when $q^{A}=q^{B}=0$, the inter-layer interdependence disappears, and the partially dependent network is simplified to two isolated networks whose dynamics are completely decoupled. In this case, faults cannot propagate between layers. In contrast, when $q^{A}=q^{B}=1$, each node is completely dependent on one or more nodes in the other layer, and the interdependent network can be regarded as a special case of the partially dependent network.

\section{Experimental Validation}
In this section, we will verify the correctness of our theoretical analysis by randomly selecting some parameters. It is not difficult to conclude from the above analysis that the robustness of the interdependent hypergraph is mainly related to three aspects: i) the probability of node survival, ii) the connectivity of the network within the layer, and iii) the dependency of the network between layers. Accordingly, we design simulations around these three dimensions, followed by empirical validation using real-world hypergraph datasets.

First, the node survival probability value range is $p=[0,1]$. In all experimental figures, we use parameter $p$ as the horizontal axis and parameter $R$ (the normalization parameter of the final network size in the steady state) as the vertical axis.

Second, we select the Poisson distribution and power-law distribution within the layer for analysis. Accordingly, we construct hypergraphs where both hyperdegree and hyperedge cardinality follow these two standard degree distributions. The relevant definitions are as follows:

1) Homogeneous hypergraph: hyperdegree and hyperedge cardinality obey Poisson distribution.
	\begin{equation}
	\begin{array}{l}
	P(k)= e^{-\langle k\rangle}\langle k\rangle^{k}/{k!},\\
	\hat{P}(m)= e^{-\langle m\rangle}\langle m\rangle^{m} /{m!}.
	\end{array}
	\label{18}
	\end{equation}

2) Heterogeneous hypergraph: hyperdegree and cardinality of hyperedges obey the power-law distribution. 
	\begin{equation}
	\begin{array}{l}
	P(k)=[(k+1)^{1-\lambda }-k^{1-\lambda }]/[{(k_{max}+1)^{1-\lambda }-k_{min}^{1-\lambda }}],\\
	\hat{P}(m)=[(m+1)^{1-\lambda }-m^{1-\lambda }]/[{(m_{max}+1)^{1-\lambda }-m_{min}^{1-\lambda }}].
	\end{array}
	\label{19}
	\end{equation}
 
Finally, we consider that the inter-layer dependency follows a Poisson distribution.
		\begin{equation}
	\tilde{P}(K)= e^{-\langle K\rangle}\langle K\rangle^{K}/{K!}.
	\label{20}
	\end{equation}

 \begin{figure}[!t]
	\centering
	\includegraphics[width=0.5\textwidth]{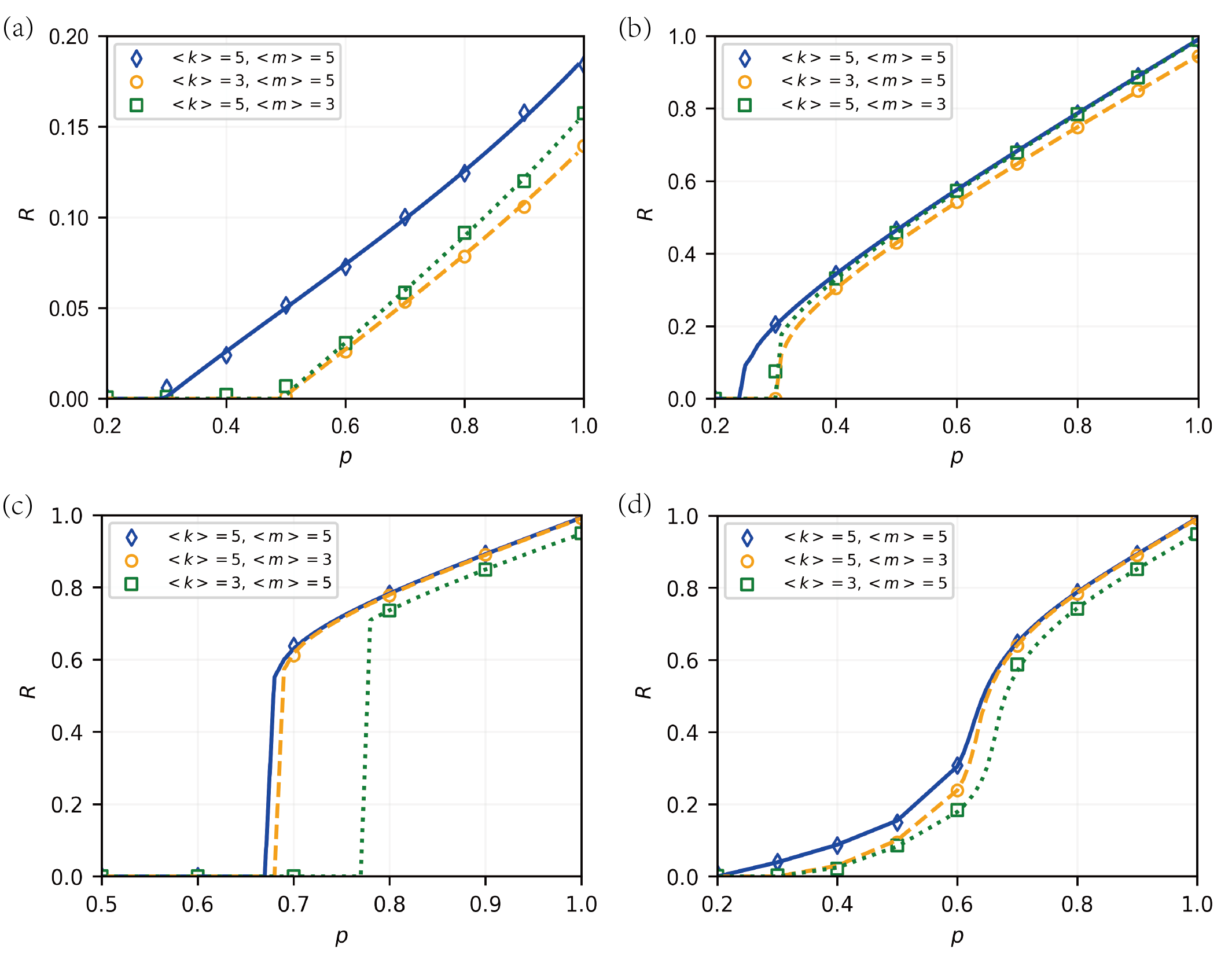}
	\caption{Effects of intra-layer parameters on robustness in interdependent homogeneous hypergraphs. (a) AND interdependence, inter-layer average degree $\left \langle K  \right \rangle=2$. (b) OR interdependence, inter-layer average degree $\left \langle K  \right \rangle=6$. (c) $\gamma$ interdependence, inter-layer average degree $\left \langle K  \right \rangle=8$ and $\gamma=0.6$. (d) $M$ interdependence, inter-layer average degree $\left \langle K  \right \rangle=8$ and $M=5$. We can find that no matter what kind of conditional interdependence, the network becomes more robust as the intra-layer parameters $\left \langle k  \right \rangle$ or $\left \langle m  \right \rangle$ increase.}
	\label{fig4}
\end{figure}

\begin{figure}[!t]
		\centering
		\includegraphics[width=0.5\textwidth]{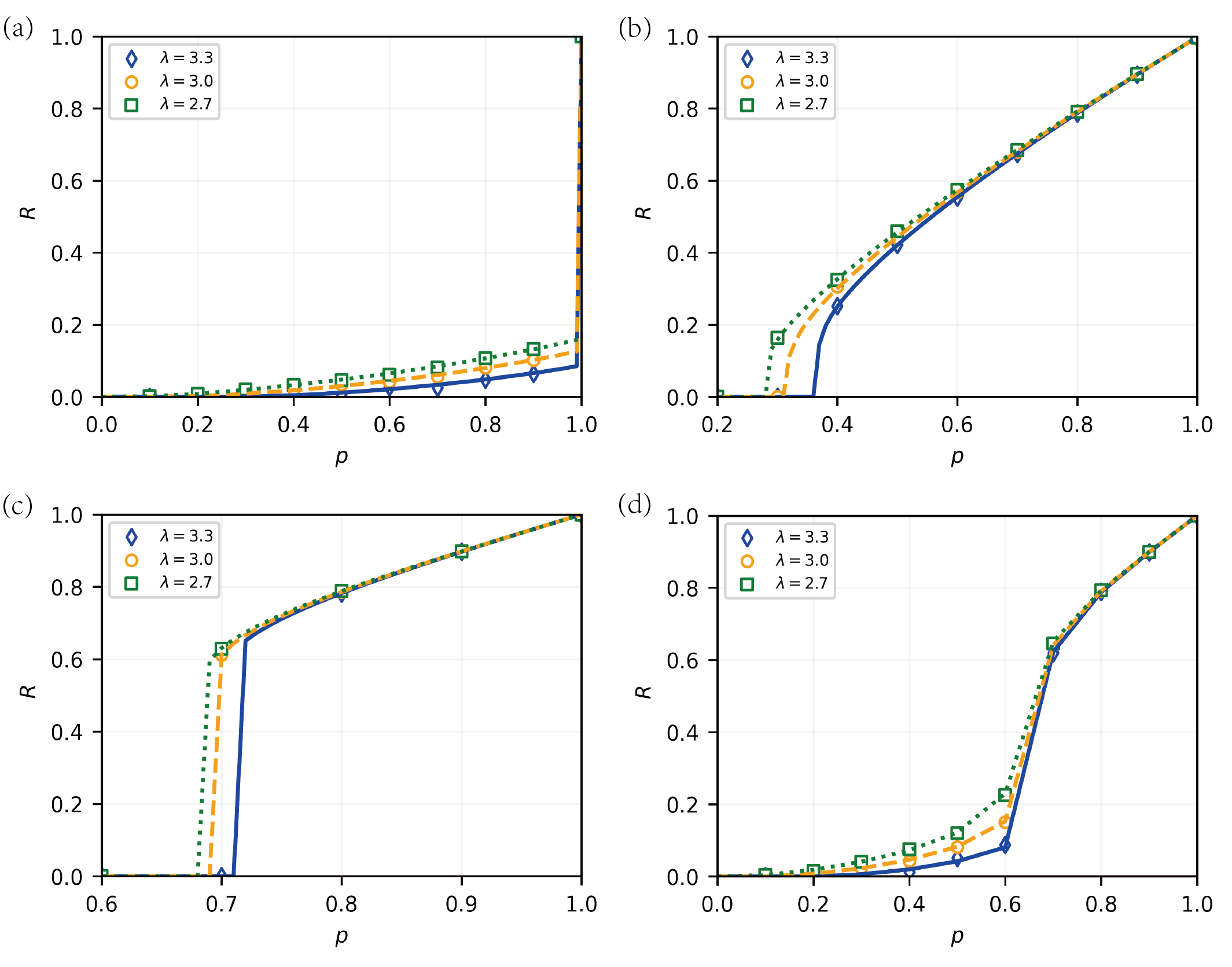}
		\caption{Effects of intra-layer parameters on robustness in interdependent heterogeneous hypergraphs. (a) AND interdependence, inter-layer average degree $\left \langle K  \right \rangle=2$. (b) OR interdependence, inter-layer average degree $\left \langle K  \right \rangle=6$. (c) $\gamma$ interdependence, inter-layer average degree $\left \langle K  \right \rangle=8$ and $\gamma=0.6$. (d) $M$ interdependence, inter-layer average degree $\left \langle K \right \rangle=8$ and $M=5$. We can find that no matter what kind of conditional interdependence, the network becomes more fragile as the intra-layer parameter $\lambda$ increase.}
	\label{fig5}
\end{figure}


\begin{figure*}[!t]
		\centering
		\includegraphics[width=1.0\textwidth]{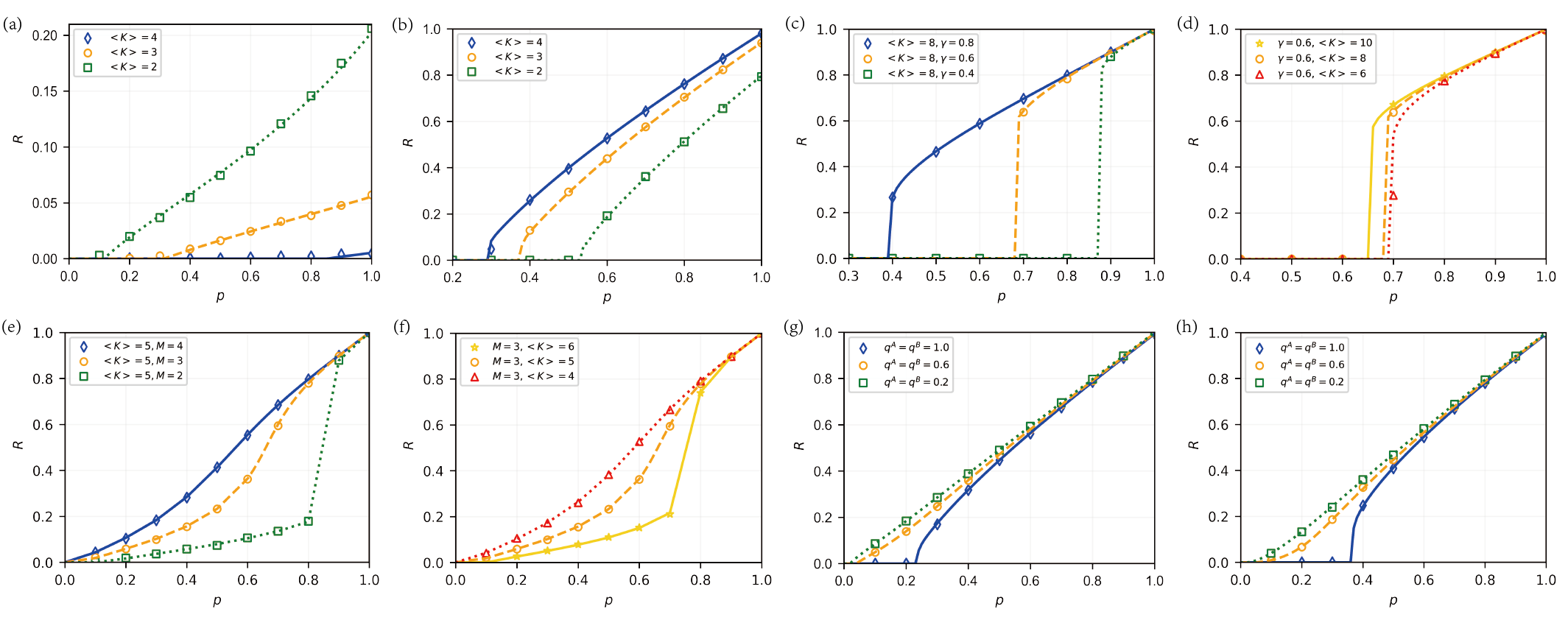}
		\caption{Effects of inter-layer parameters on robustness in interdependent hypergraphs. (a)-(g) are the results of experiments conducted under interdependent homogeneous hypergraphs ($\left \langle k \right \rangle=\left \langle m \right \rangle=8$). (a) AND interdependence, as $\left \langle K \right \rangle$ increases, the network robustness becomes fragile. (b) OR interdependence, as $\left \langle K  \right \rangle$ increases, the network becomes more robust. (c) and (d) $\gamma$ interdependence, (c) fixed $\left \langle K \right \rangle$ = 8, as $\gamma$ increases, the network becomes more robust; (d) fixed $\gamma$ = 0.6, as $\left \langle K \right \rangle$ increases, the network becomes more robust. (e) and (f) $M$ interdependence, (e) fixed $\left \langle K \right \rangle$ = 5, as $M$ increases, the network becomes more robust; (f) fixed $M$ = 3, as $\left \langle K \right \rangle$ increases, the network becomes more robust.} (g) Partial dependence under OR interdependence, as the coupling strength increases, the network robustness becomes more fragile. And (h) Partial dependence under interdependent heterogeneous hypergraphs ($\lambda=3.0$ and $\left \langle K \right \rangle=5$), the conclusion is consistent with the conclusion in the homogeneous hypergraph.
	\label{fig6}
\end{figure*}

\subsection{Influence of intra-layer parameters} 
Next, we conducted simulation experiments. In the case of homogeneous hypergraphs, the main parameters within the layer are $\left \langle k  \right \rangle$ and $\left \langle m \right \rangle$. To explore the impact of different parameters on network robustness, we coupled two homogeneous hypergraphs with $N$=10,000 nodes, fixed the average number of dependent edges between layers to $\left \langle K  \right \rangle$, and formed a one-to-many interdependent homogeneous hypergraph network by changing the values of parameters $\left \langle k  \right \rangle$ and $\left \langle m \right \rangle$. On this basis, we randomly removed nodes with a ratio of $(1-p)$ in hypergraphs $A$ and $B$ respectively, and conducted 100 independent repeated experiments according to the four inter-layer dependency conditions, and calculated the average of the experimental results as a measure of the steady-state scale of the network under the final dependency conditions. We represent the simulation results with symbols, the theoretical analysis results with lines, and the relevant experimental results are shown in Fig. \ref{fig4}. It is not difficult to see that as $\left \langle k \right \rangle$ or $\left \langle m \right \rangle$ increases, the connections between nodes in the layer will become tighter, making the network more robust.

Similar to the experimental setting on the homogeneous hypergraph, the main parameter in the heterogeneous hypergraph is the power exponent $\lambda$. We couple two heterogeneous hypergraphs with $N$=10,000 nodes, fix the average number of dependency edges between layers to $\left \langle K  \right \rangle$, and change the value of the parameter $\lambda$ to form a one-to-many interdependent heterogeneous hypergraph network. From the results shown in Fig.\ref{fig5}, it can be seen that as the power exponent $\lambda$ increases, the network becomes more fragile.

In addition, according to the results in Fig.\ref{fig4} and Fig.\ref{fig5}, as the probability of a node being removed increases, the network will become more fragile. We also notice that in the AND interdependence in Fig.\ref{fig5} (a), the initial interdependent heterogeneous hypergraph network $(p=1)$ will collapse on a large scale after a slight disturbance. This also shows that the conditional dependence of AND interdependence is too strict for node survival.

\subsection{Influence of inter-layer parameters}	
For inter-layer interdependency, different conditional interdependencies have different inter-layer parameters. Therefore, we conduct experiments in various situations. For AND interdependence, the main parameter is the inter-layer average degree $\left \langle K  \right \rangle$. We experimented with different values, and the results are shown in Fig.\ref{fig6} (a). As the value of $\left \langle K \right \rangle$ increases, the network becomes more fragile. This is because the rise in $\left \langle K  \right \rangle$ will increase the nodes' dependency edges, and the conditions for node survival become more stringent. It is not difficult to see that when $\left \langle K  \right \rangle \to \infty $, the network will always collapse.	 

For OR interdependence, consistent with AND interdependence, its main parameter is the average degree $\left \langle K  \right \rangle$ between layers. We experimented with different values, and the results are shown in Fig.\ref{fig6} (b). Unlike the conclusion of AND interdependence, as the value of $\left \langle K  \right \rangle$ increases, the network becomes more robust. This is because the rise in $\left \langle K  \right \rangle$ will increase nodes' dependency edges, and the probability of node survival will also improve significantly. It is not difficult to see that when $\left \langle K  \right \rangle \to \infty $, the interdependent network is similar to two independent networks.

The main inter-layer parameters in the $\gamma$ interdependence are $\left \langle K  \right \rangle$ and $\gamma$. To this end, we fixed$\left \langle K  \right \rangle$ and $\gamma$ and conducted multiple experiments. As shown in Figs. \ref{fig6} (c) and (d), we found that when $\left \langle K \right \rangle=8$ is fixed, the network becomes more robust as $\gamma$ increases; when $\gamma=0.6$ is fixed, the network becomes more robust as $\left \langle K \right \rangle$ increases. This is because in the equation $\left \lfloor  K \gamma \right \rfloor$, as $\left \langle K  \right \rangle$ or $\gamma$ increases, the number of dependent edges allowed to fail also increases, thereby improving the robustness of the interdependent hypergraph.

Similarly, the inter-layer parameters involved in the $M$ interdependence are mainly $\left \langle K  \right \rangle$ and $M$. To this end, we fixed $\left \langle K  \right \rangle$ and $M$ and conducted multiple experiments. As shown in Figs. \ref{fig6} (e) and (f), we found that when $\left \langle K \right \rangle=5$ is fixed, as $M$ increases, the network becomes more robust; when $M=3$ is fixed, as $\left \langle K \right \rangle$ increases, the network becomes more fragile. This is because when the ratio $M / \left\langle K \right\rangle$ becomes larger, the number of dependent edges allowed to fail also increases, thereby improving the robustness of the interdependent hypergraph.

The similarities and differences between $\gamma$ interdependence and $M$ interdependence lie in the fact that when each node has the same fixed number of dependent edges $K$ (i.e., $\langle K \rangle = K$), the two interdependence mechanisms become equivalent under the condition $\gamma = M/K$, and can thus be transformed into each other. However, such an ideal scenario is rarely encountered in practice. In $\gamma$ interdependence, when $\gamma \in [0, 1)$, each node may fail if it does not satisfy the dependency condition; when $\gamma = 1$, the survival of all nodes is not affected by the state of their dependent edges. In contrast, under $M$ interdependence, only those nodes with $K \leq M$ are not affected, regardless of the failure state of their dependent edges.

The above is the effect of inter-layer parameters on network robustness under four conditional interdependence. In OR interdependence, the condition that at least one dependency edge is required for the survival of inter-layer nodes is too strict.

Therefore, we next consider hypergraph networks with partial dependencies in OR interdependence. We generate two homogeneous hypergraphs $A$ and $B$, where the parameters $ \left \langle k \right \rangle  =   \left \langle m \right \rangle= 8$ , and $\left \langle K  \right \rangle=5$ in each hypergraph. For simplicity, we set the inter-layer coupling strength $q^{A}=q^{B}$, and we conduct multiple groups of experiments by changing the coupling strength. As shown in Fig.\ref{fig6} (g), we found that when the inter-layer coupling strength is small, the failure of a node in one layer has little impact on the other layer. As the inter-layer coupling strength increases, the connection between the hypergraphs becomes closer, and the range of damage propagation between the networks will also be larger, resulting in a continuous decrease in the scale of the final network in steady state.	

We also generate two heterogeneous hypergraphs $A$ and $B$, where the parameter $\lambda$ = 3 and $\left \langle K \right \rangle=5$ in each hypergraph. The same experiment is performed and the results are shown in Fig.\ref{fig6} (h). The conclusions are consistent with those in the homogeneous hypergraph.

\subsection{Validation on Empirical Hypergraph Datasets} 
To date, network datasets with both higher-order hyperedge structures and cross-layer interdependencies are still scarce. Existing hypergraph data usually only represent single-layer networks, such as co-authorship networks and contact networks, and do not contain explicit dependency mappings between different layers. To this end, this paper constructs an interdependent higher-order network system, in which: each layer is derived from a real hypergraph dataset, layer $A$ is a Primary school temporal network (Primary school) \cite{stehle2011high,gemmetto2014mitigation}, layer $B$ is a High school contact and friendship networks (High school) \cite{mastrandrea2015contact}. By combining two real-world hypergraph datasets with compatible structures and semantics, we construct interdependent higher-order networks in a scientifically accepted manner, detailed parameters are shown in Table \ref{biao}. The one-to-many dependency relationship between nodes in the layers is generated by random mapping, where the nodes in layer $A$ depend on multiple nodes in layer $B$, and the nodes in layer $B$ also depend on multiple nodes in layer $A$, with an average number of dependency edges of $\left \langle K \right \rangle$. This construction provides us with an experimental environment that is both realistic and easy to control for evaluating the robustness of such systems in the face of cascading failures.

\begin{table}[htbp]
\centering
\caption{Statistics of Real-World Hypergraph Datasets}
\small
\setlength{\tabcolsep}{8pt}
\renewcommand{\arraystretch}{1.2}
\begin{tabular}{lcccccc}
\toprule
\textbf{Dataset} & $N$ & $E$ & $\langle k \rangle$ & $\langle m \rangle$\\
\midrule
Primary school &   242    &   1917    &  24.1297     &  3.0454  \\
High school &   327    &    1453   &  12.8104     &   2.8830   \\
\bottomrule
\end{tabular}
\vspace{0.5em}
\begin{flushleft}
\scriptsize
Note: $N$ = number of nodes; $E$ = number of hyperedges; $\langle k \rangle$ = average hyperdegree; $\langle m \rangle$ = average hyperedge size.
\label{biao}
\end{flushleft}
\end{table}

The nodes in each layer of the hypergraph are students in the school. Wearable sensors are used to capture close contact between students. The contact information is aggregated into a 20-second time window, and each hyperedge is the maximal clique within each contact interval \cite{peng2022targeting}. In this set of experiments, the intra-layer topology of each real-world hypergraph remains fixed. We vary the inter-layer dependency strength to evaluate the applicability and validity of the theoretical analysis under different coupling scenarios. 

To quantitatively compare the consistency between theoretical and simulation results, we calculated the error under different parameter settings. Specifically, for each parameter $p_i \in \{0.1, 0.2, \dots, 1.0\}$, let $R_{\text{theory},i}$ and $R_{\text{sim},i}$ 
denote the theoretical and simulation values, respectively. We use the absolute error to avoid division by zero 
\begin{equation}
e_i = \left| R_{\text{sim},i} - R_{\text{theory},i} \right| \times 100\% .
\label{21}
\end{equation}

All $e_i$ for $p_i \in \{0.1,0.2,\dots,1.0\}$ are then averaged to obtain the mean absolute error $\overline{e}$, which serves as an overall measure of the agreement between theory and simulation. The corresponding results are presented in Fig.\ref{fig7}. As observed, the theoretical results exhibit excellent agreement with the simulations on interdependent hypergraphs constructed from real-world datasets, thereby validating the effectiveness of the proposed analytical framework.

 \begin{figure}[!t]
\centering
\includegraphics[width=0.5\textwidth]{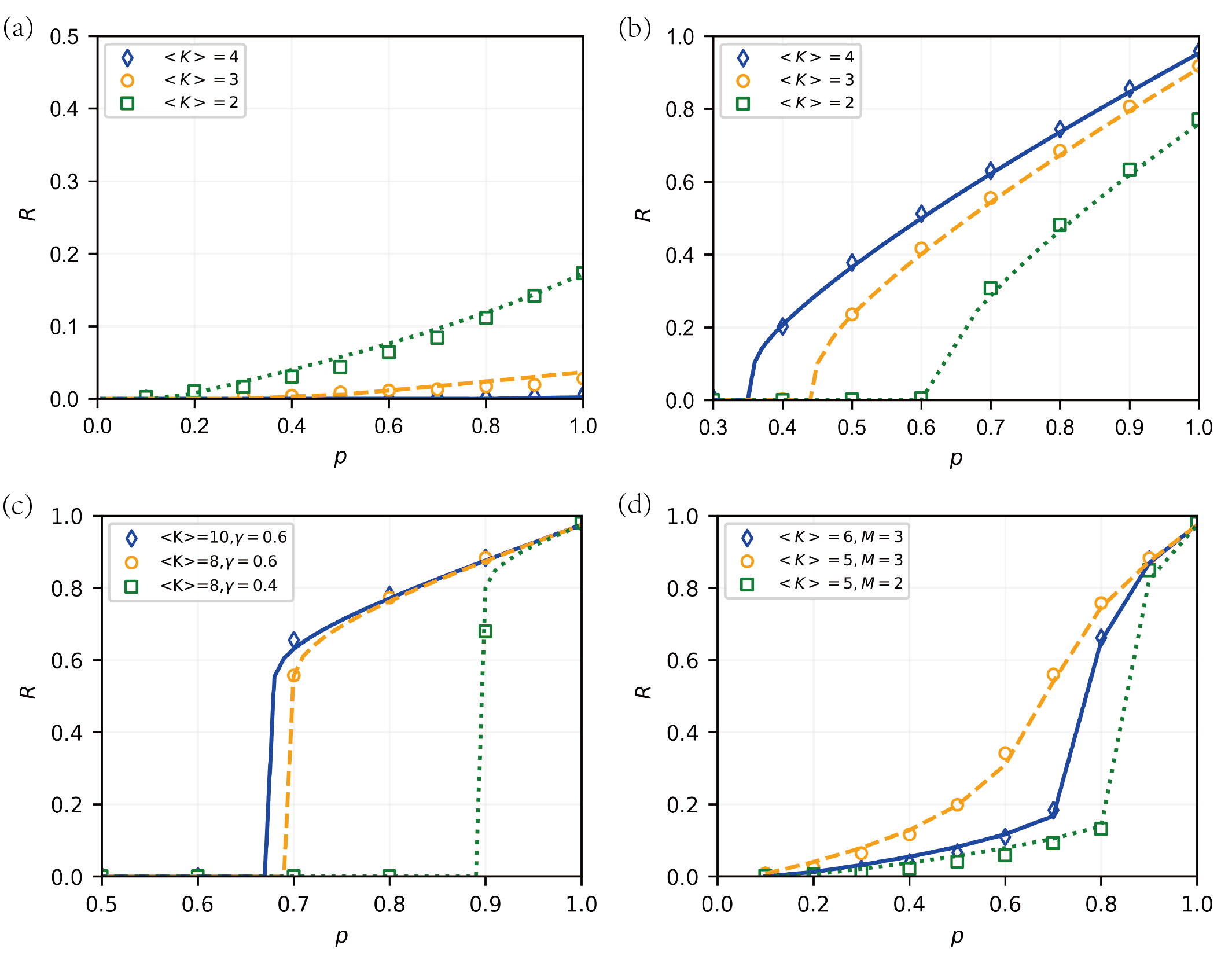}
\caption{Robustness of interdependency hypergraphs constructed based on real-world data. (a) AND interdependence: the average absolute errors for $\left \langle K \right \rangle=2,3,4$ are 0.69\%, 0.41\%, and 0.16\%, respectively. (b) OR interdependence: the average absolute errors for $\left \langle K \right \rangle=2,3,4$ are 0.72\%, 0.68\%, and 0.73\%, respectively. (c) $\gamma$ interdependence: the average absolute errors for $(\left \langle K \right \rangle=8,\gamma=0.4)$, $(\left \langle K \right \rangle=8,\gamma=0.6)$, and $(\left \langle K \right \rangle=10,\gamma=0.6)$ are $0.18\%$, $0.29\%$, and $0.50\%$, respectively. (d) $M$ interdependence: the average absolute errors for $(\left \langle K \right \rangle=5,M=2)$, $(\left \langle K \right \rangle=5,M=3)$, and $(\left \langle K \right \rangle=6,M=3)$ are $1.17\%$, $1.26\%$, and $0.98\%$, respectively. The solid lines represent the theoretical predictions, while the symbols denote the average simulation results over 100 independent runs. The high consistency between the theoretical analysis and simulation results demonstrates the effectiveness of the proposed model.}
\label{fig7}
\end{figure}

\section{CONCLUSION} 
This paper uses two hypergraphs to form an interdependent hypergraph with one-to-many dependencies. Based on this, we propose a unified framework to study four different inter-layer dependency conditions. This not only extends our previous research on failure cascades on one-to-one interdependent hypergraphs, but also further expands the previous related research on inter-layer dependencies in pairwise interaction networks.

Interdependent networks are usually composed of different types of edges, with connectivity within the layer and dependencies between layers. When the network is subjected to attacks, these edges become different paths for damage propagation. Under the joint action of edges with different properties, cascading failures inevitably profoundly impact the dynamics of the entire network. Against this background, we propose a theoretical framework to analyze the robustness of such systems. Our theoretical analysis performs well on interdependent homogeneous and heterogeneous hypergraphs, as well as on interdependent networks derived from real-world hypergraph data. Through the experimental results, we draw the following conclusions: To improve the robustness of the network, first, we can improve the node survival probability \( p \); second, we can improve the robustness by enhancing the intra-layer connectivity; for homogeneous hypergraphs, we can improve by increasing $\left \langle k \right \rangle$ or $\left \langle m \right \rangle$; for heterogeneous hypergraphs, reducing \( \lambda \) helps to enhance the network robustness. These results suggest that the structure of higher-order interactions plays an important role in determining network robustness. In particular, hyperedges with larger cardinalities can better maintain network connectivity in the presence of node failures, due to the increased redundancy in multi-node interactions; finally, according to different inter-layer dependency conditions, corresponding strategies are adopted: $\left \langle K \right \rangle$ should be reduced under AND dependency conditions, $\left \langle K \right \rangle$ should be increased under OR dependency conditions, $\left \lfloor  K \gamma \right \rfloor$ should be increased under $M$ dependency conditions, and $M/\left \langle k \right \rangle $ should be increased under $M$ dependency conditions. In addition, this paper also considers the impact of coupling strength on network robustness. Experiments show that reducing the coupling strength between networks can also improve the robustness of the network.

As the network topology becomes increasingly complex, the research results of this paper have essential scientific guiding significance for improving the robustness of interdependent systems in the real world, especially in complex network environments, providing system designers with specific optimization strategies for different types of dependencies. These studies not only help to enhance the robustness and fault tolerance of large-scale networks but also provide theoretical basis and practical guidance for dealing with problems such as network interruption and information dissemination failure. This paper considers the case where a hyperedge is considered failed only when all nodes fail. However, in the context of hyperedge failure caused by the failure of only some nodes \cite{bianconi2024nature,bianconi2024theory}, developing a one-to-many interdependent hypergraph network model is a direction worth further exploration.}

With the integration of cross-domain technologies, future research can further explore the interaction between multi-network systems, such as the coupling effect between the Internet of Things and intelligent transportation, energy systems and communication networks. In this process, optimising intra-layer connectivity, inter-layer dependencies and node survival rate will be essential for improving system reliability. Furthermore, consideration can be given to studying adaptive dependency mechanisms that enable network nodes to dynamically switch dependencies, thereby enhancing the flexibility and robustness of multi-layer systems. Combining theoretical models with practical applications and promoting the verification and optimization of interdependent networks can provide new perspectives and useful methods for the reliability research of complex systems. In addition, future work can also combine machine learning and data-driven technologies to improve network modeling and analysis further and improve network fault prediction, recovery capabilities and optimal configuration strategies, thereby further improving the overall robustness and adaptability of the system.

\bibliographystyle{ieeetr}
\bibliography{mybibtex}

\begin{IEEEbiography}[{\includegraphics[width=1in,height=1.25in,clip,keepaspectratio]{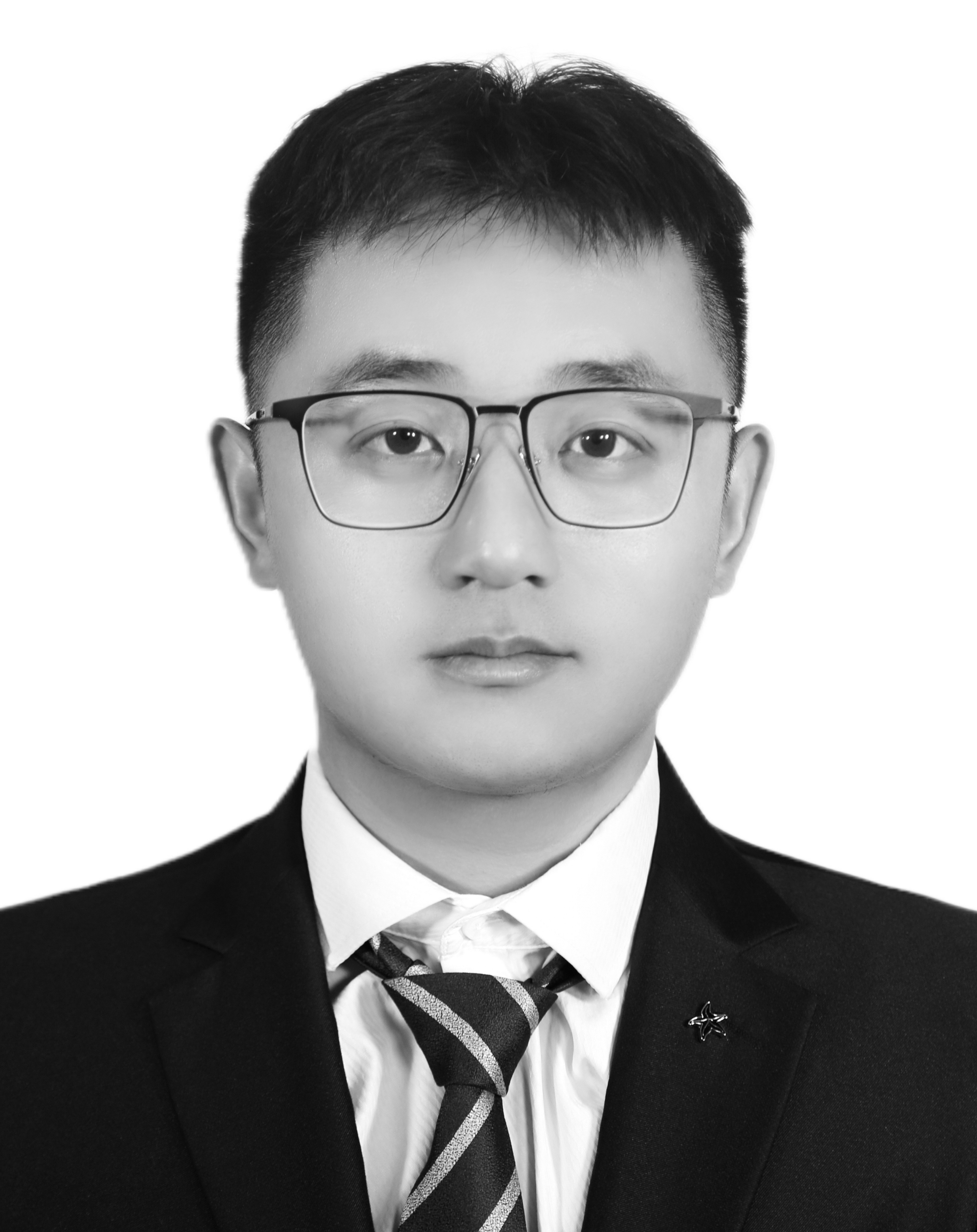}}]{Cheng Qian} is currently a PhD student (pursuing a combined master's and PhD program) at the School of Computer Science and Technology, Zhejiang Normal University, Jinhua, Zhejiang 321004. His research focuses on theoretical studies in network science and robustness analysis of higher-order systems. He has published over ten papers in the field of network science.
\end{IEEEbiography}

\begin{IEEEbiography}[{\includegraphics[width=1in,height=1.25in,clip,keepaspectratio]{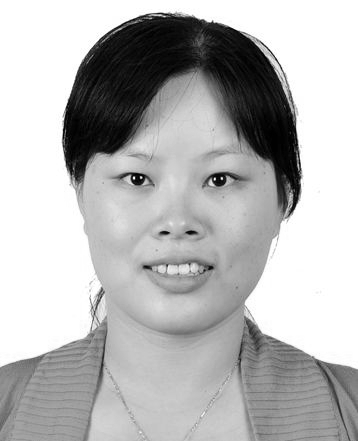}}]{Dandan Zhao} is currently an associate professor of the School of Computer Science and Technology of Zhejiang Normal University. She received her Ph.D degree in circuits and systems from Shanghai Jiaotong University.  Her current research interests include Cyber Physical System, Network Security and Graph Theory.
\end{IEEEbiography}

\begin{IEEEbiography}[{\includegraphics[width=1in,height=1.25in,clip,keepaspectratio]{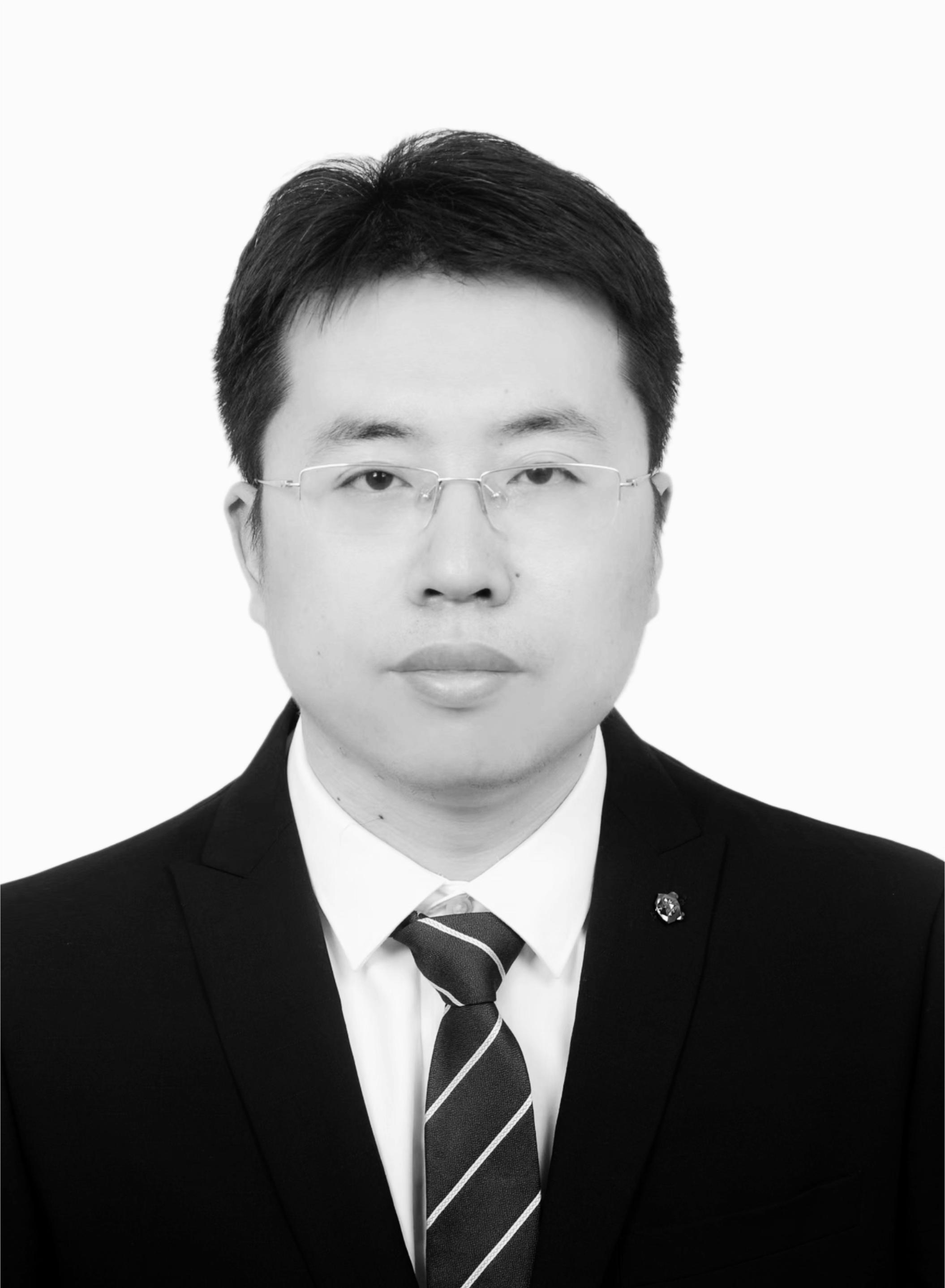}}]{Bo Zhang} received the B.S. degree from Hohai University, the M.Sc. degree from Southeast University, and the PhD from Nanjing university of science \& technology. He is currently a postdoc at shanghai jiao tong university, China. His research interests include cybersecurity in smart grid and network security situation awareness.
\end{IEEEbiography}

\begin{IEEEbiography}[{\includegraphics[width=1in,height=1.25in,clip,keepaspectratio]{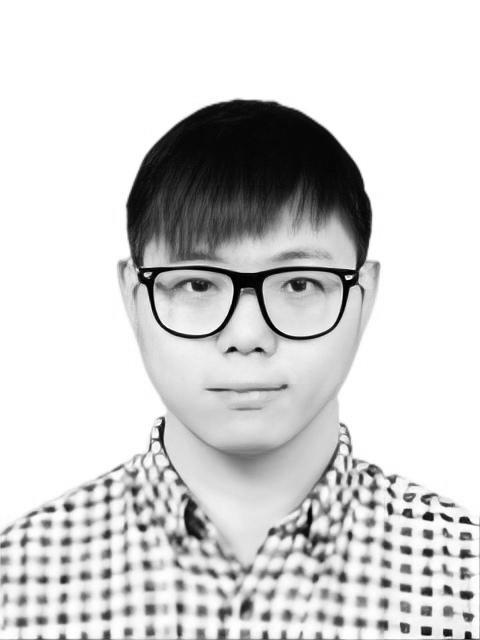}}]{Ming Zhong} is currently a lecturer in the School of Computer Science and Technology of Zhejiang Normal. He is also an enterprise postdoc at Zhejiang Dingren Fire Technology. He received his Ph.D degree in computer science from Zhejiang University. He was a visiting scholar at Hongkong City University and a researcher at NetEase. His current research interests include Network Security, AI Security and Data Analysis.
\end{IEEEbiography}

\begin{IEEEbiography}[{\includegraphics[width=1in,height=1.25in,clip,keepaspectratio]{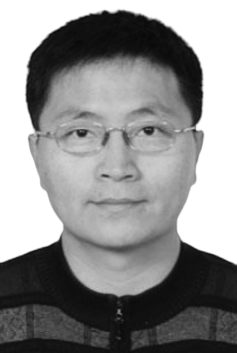}}]{Jianmin Han} received the BS degree from the Department of Computer Science and Technology, DaQing Petroleum Institute, in 1992, and the PhD degree from the Department of Computer Science and Technology, East China University of Science and Technology, in 2009. He is currently a professor with the College of Computer Science and Technology, Zhejiang Normal University. His research interests include privacy preservation and game theory.
\end{IEEEbiography}

\begin{IEEEbiography}[{\includegraphics[width=1in,height=1.25in,clip,keepaspectratio]{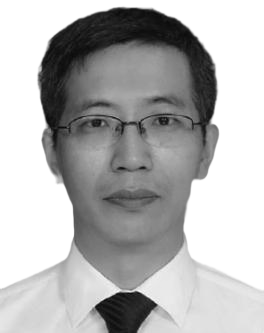}}]{Shenghong Li (Senior Member, IEEE)}  received the Ph.D. degree in radio engineering from the Beijing University of Posts and Telecommunications, Beijing, China, in 1999. He is currently a Professor with the School of Electronic Information and Electrical Engineering, Shanghai Jiao Tong University, Shanghai, China. His current research interests include network and information security, signal and information processing, and machine learning. He has authored or co-authored over 100 papers, co-authored 4 books, and holds over 20 granted patents. Dr. Li was a recipient of the 1st Prize of Shanghai Science and Technology Progress in China in 2003 and 2013. In 2006 and 2007, he was elected for New Century Talent of the Chinese Education Ministry and Shanghai Dawn Scholar.
\end{IEEEbiography}

\begin{IEEEbiography}[{\includegraphics[width=1in,height=1.25in,clip,keepaspectratio]{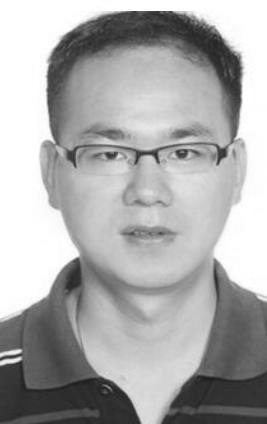}}]{Hao Peng (Member, IEEE)} is currently a professor in the School of Computer Science and Technology of Zhejiang Normal University. He received his Ph.D degree in communication and information system from Shanghai Jiaotong University. He serves as executive dean of AI Research Academy, Director of Network Security \& Optimization Research Institute and Head of Computer Science Department in Zhejiang Normal University. He is sponsored by Special Support Program for High-level Talents in Zhejiang Province and National Natural Science Foundation of China. His current research interests include Network Security, AI Security and Data-driven Security. He is a member of IEEE, CAAI and CCF.
\end{IEEEbiography}

\begin{IEEEbiography}[{\includegraphics[width=1in,height=1.25in,clip,keepaspectratio]{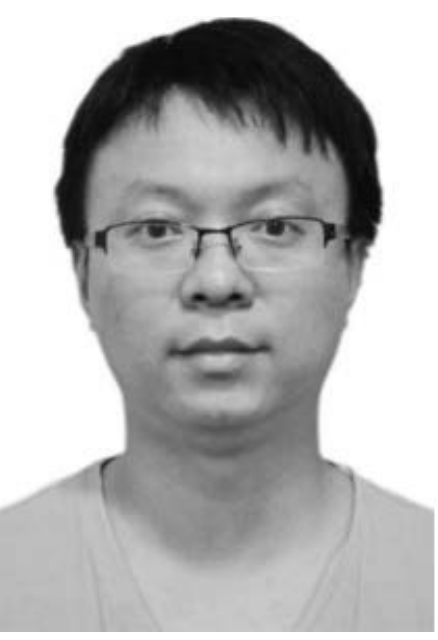}}] {Wei Wang }received the Ph.D. degree from the University of Electronic Science and Technology of China, Chengdu, China, in 2017. He is currently a Professor at Chongqing Medical University, Chongqing. He has published over 100 research articles on network science and spreading dynamics.

\end{IEEEbiography}
\end{document}